\documentclass[11pt,a4]{article}
\usepackage{theorem}
\usepackage{url}
\usepackage{indentfirst}
\usepackage{graphicx}

\usepackage[labelformat=simple]{subcaption}

\usepackage{hyperref}
\hypersetup{
	colorlinks = true,
	citecolor = blue,
}

\emergencystretch=\maxdimen
\begin{document}

\title{Feedforward Neural Networks for Caching: \\Enough or Too Much?}

%

\author{
Vladyslav Fedchenko \\
Universit\'e C\^ote d'Azur, Inria\\ 
Sophia Antipolis, France\\
\texttt{vladfedchenko@gmail.com}
\and
Giovanni Neglia \\
Universit\'e C\^ote d'Azur, Inria\\ 
Sophia Antipolis, France\\
       \texttt{giovanni.neglia@inria.fr}
\and
Bruno Ribeiro\footnote{Authors are listed in alphabetical order.} \\
Purdue University\\
West Lafayette, IN, USA\\
       \texttt{ribeiro@cs.purdue.edu}
}


\maketitle

\begin{abstract}
We propose a caching policy that uses  a feedforward neural network (FNN) to predict content popularity. Our scheme outperforms popular eviction policies like LRU or ARC, but also a new policy relying on the more complex recurrent neural networks. At the same time,  replacing the FNN predictor with a naive linear estimator does not degrade caching performance significantly, questioning then the role of neural networks for these applications.
\end{abstract}




\section{Introduction}
Caching is doubly beneficial: it reduces data retrieval time by storing a copy at a closer/faster-accessible location and, at the same time, decreases the load on the remote system where the original version is located.
For this reason, caches are ubiquitous in IT systems, ranging from L2 caches built into the CPU to in-memory page caches managed by the operating systems, from local web proxy caches to Internet-wide Content Delivery Networks (CDNs) and cloud-based in-memory key-value stores like Amazon's ElastiCache.

It is difficult to decide which contents should be stored in the cache and which ones should be evicted. Even when the sequence of future requests is known in advance, maximizing the hit ratio is in general a strongly NP-complete problem~\cite{chrobak12}. Moreover, in most cases of practical interest, future requests are unknown and a caching policy may only try to guess what will happen. To this purpose, the policy usually looks at the past sequence of requests to exploit possible elements of predictability, like the fact that future requests are often more likely to be for recently referenced contents (temporal locality) or for related ones (spatial locality). But additional information could be beneficial. For example, in a CDN the time of the day, users' profiles, and information about what is happening at close-by caches are likely to be correlated with future requests.

The first paper to propose the idea to use machine learning to learn caching rules from available rich data was probably~\cite{bastug15}. Nevertheless, the only practical example considered there was collaborative filtering to estimate content popularities at some locations from measurements at other locations. 
Surprisingly, the idea to use neural networks (NNs) for caching purposes was only explored during the last year in~\cite{zeng17,hashemi18,tsai18,narayanan18,zhang18}. All these papers (described  in Sect.~\ref{s:related}) adopt recurrent neural networks with long short-term memory  units (LSTM in what follows), motivated by the fact that LSTM have proved to be very effective to address sequence prediction problems such as those found in natural language processing. 

LSTM neural networks (LSTM-NNs), as all recurrent networks,  have  feedback loops which give them some kind of memory. They are then more complex (and more difficult to train) than the classic feed-forward NNs (FNNs) where signals can only travel one way from the input layer to the output one. The question at the origin of our work was then if the simpler FNNs could also perform well for caching purposes. Answering this simple question has lead us to unexpected conclusions.

The paper is organized as follows. After an overview of the related work in Sect.~\ref{s:related}, we describe our caching policy and its FNN predictor for content popularity in Sect.~\ref{s:policy}. In Sect.~\ref{s:peva} we present our performance evaluation results on real traces from Akamai CDN. Sect.~\ref{s:discussion} concludes the paper.

\section{Related work}
\label{s:related}

Papers~\cite{zeng17,hashemi18} propose two different LSTM-NNs for prefetching inside a processor. Instead, in our paper, we  focus on internet applications like web proxies, or content delivery networks (CDNs), and we consider reactive caching, i.e.~contents can be retrieved only once they are requested. Moreover, \cite{zeng17,hashemi18} evaluate the quality of their solutions only in terms of prediction accuracy and not of the final data retrieval time or other system metrics prefetching should improve. In this paper we conclude that the relation between prediction accuracy and final performance of the caching system may be weaker than one would expect.

The authors of \cite{tsai18} study caching at the base stations of a cellular network. In their case the LSTM-NN is used to perform a sentiment analysis of mobile users' traffic to detect their interests and prefetch  contents they may like.

Reference~\cite{narayanan18} proposes to use an LSTM-NN as an additional module to be integrated with an existing baseline caching policy like LRU. The LSTM-NN generates fake requests that are added to the real request stream. These additional requests push the baseline caching policy to prefetch or maintain in the cache contents that are likely  to be used in the future.

Finally, \cite{pang18} is the closest paper to ours, because it  uses an LSTM-NN as a popularity predictor and then manages the cache as a priority queue where, upon a miss,  contents with the smallest predicted popularity are evicted. We have implemented the caching policy in \cite{pang18} and compared it with ours in Sect.~\ref{s:peva}.

Another set of papers like~\cite{zhong18,rezaei18} use reinforcement learning to adapt dynamically the caching policy. 
A reinforcement learning algorithm  tries to learn directly the best action to pursue, often by a trial-and-error approach. For this reason, training a reinforcement learning algorithm is often more complex than learning the parameters of a neural network, and results may be unstable~\cite{henderson18}.

\begin{table}
\center
\begin{tabular}{| c | p{4.5cm} | c | }
\hline
Symbol &  Meaning & Typical Value\\
\hline
$T$	& epoch duration & 200 s \\
$K$ & \# of epochs &  4\\
$c$	& constant in the logarithmic mapping & $10^{-15}$\\
$\alpha$ & leaky ReLU coefficient & $10^{-2}$\\
$\eta$ & learning rate & $10^{-4}$ \\
$\gamma$ & learning rate discount & 0.5 \\
$H$		& \# of old epochs used for training & 9\\
\hline
\end{tabular}
\caption{Notation and typical values.}
\label{t:notation}
\end{table}

\section{Caching policy}
\label{s:policy}
The core idea of our policy is simple and was naturally adopted by previous works like~\cite{li16} and~\cite{pang18}: we keep in the cache the contents with the largest estimated popularities. The popularity of a content is defined as the fraction of requests for that content over a meaningful time horizon. When contents have the same size (as we assume) this strategy should lead to maximize the cache hit ratio.

\subsection{Components}
In detail, our caching system is composed by a feature database,  a popularity predictor, and a content storage element. 

The content storage is managed through a min-heap data structure with the estimated popularities as keys. In this way, it is possible to retrieve fast a pointer to the content with the minimum estimated popularity.

The feature database contains the current feature vector for each of the contents in the catalogue.\footnote{
	This solution has clearly a large memory footprint, 
	but  in this paper we want first to evaluate the potentialities of NN-based caching.
} The feature vector  has $K$ elements: it contains the fractions of requests (popularities) for  content $i$ during the previous $K-1$ epochs ($p_{-(K-1),i}, \dots p_{-1,i}$) and during the current one ($p_{0,i}$). Each epoch has the same duration $T$. We remark that, during a given epoch, past popularities do not change, but current popularity does. Table~\ref{t:notation} lists the main parameters of our caching policy together with their typical values used for the experiments in~Sect.~\ref{s:peva}.

The popularity predictor receives as input 1) the feature vector of a content and 2) how much time $t$ has passed since the beginning of the current epoch ($0\le t \le T$). It outputs the popularity estimation for the current epoch.\footnote{
	It could be more useful to estimate the popularity during a time interval of duration $T$ in the future, but this choice simplifies the learning phase.
}
The prediction algorithm is based on an FNN. It is the focus of this paper and is described in detail in~Sect.~\ref{s:fnn_popularity}.

\subsection{Operation}
The caching policy works as follows. When a  request for content $i$ arrives, the feature vector of content $i$ is updated and provided as input to the popularity predictor together with the time $t$. The  predictor estimates then  the current popularity of content $i$ ($\hat p_{0,i}$).

If content $i$ is present in the cache (a  hit occurs), it is served and its popularity key is updated in the heap. If it is not stored locally (a miss occurs), it is retrieved from the authoritative server or from a higher-level cache and served. Moreover, the policy decides if storing it locally. To this purpose, the policy compares  the popularity estimate $\hat p_{0,i}$ with the minimum popularity of the contents currently in the cache. If $\hat p_{0,i}$ is larger, then content $i$ replaces the least popular content, otherwise, it is discarded. 

Moreover, upon a request, 
a few contents in the cache are randomly selected and their popularity is re-evaluated through the predictor. This prevents old contents, which are no more requested, but whose popularity was overestimated in the past, from staying forever in the cache.

At the end of each epoch, the predictor is updated through a new training phase, described below.

\subsection{Popularity predictor}
\label{s:fnn_popularity}
The core of the predictor is a Feedforward Neural Network (FNN).
The FNN has $K+1$ input values 
 and one output value. The input values are the time $t$ since the beginning of the current epoch and a mapping of the feature vector of $K$ popularities ($x_{-l,i} = F(p_{-l,i})$), where $F(p) = -\log(p + c)$ and $c>0$ is a small constant. This transformation makes the input vector more homogeneous (popularities differ by many order of magnitudes). The output $y_i$ of the FNN is interpreted as the mapping of the current popularity. The final predicted popularity is then obtained through the inverse transformation $\hat p_{0,i} = F^{-1}(y_i)$.

Internally, the FNN has 2 fully connected hidden layers, each with 128 neurons. Each neuron uses a rectified linear unit (ReLU) as activation function ($a(x)$), and in particular  a Leaky ReLU   to overcome the ``dying ReLU'' problem \cite{14}. Then $a(x)=x$, if $x\ge 0$, and $a(x)= \alpha x$ otherwise, where $\alpha$ is a small constant. In our experiments, Leaky ReLU provides better results than the usual sigmoid.

The FNN is trained at the end of each epoch using the contents requested during the epoch. 
The  prediction quality is evaluated through the squared difference $(y_i - F(p_{0,i}))^2$ (note that at the end of one epoch  $p_{0,i}$ is the correct popularity of content $i$). The loss function is then the classic mean squared error (MSE). 
FNN weights are updated through the  usual backpropagation mechanism with learning rate $\eta$. 

Moreover, the  training datasets for the previous  $H$ epochs are reused, but with a discounted learning rate $\gamma^i \eta$ for the $i$th epoch farthest in the past. This is a standard technique to prevent the problem of catastrophic forgetting~\cite{16, 17} that may occur when the training uses only the most recent data. In this case, the neural network can indeed forget the information about some underlying relations between input and output learned in the past, even though they still may be relevant for predictions.

\section{Performance Evaluation}
\label{s:peva}
We have evaluated the performance of our caching policy both on synthetic and on real traces. 

In the synthetic trace, requests for contents arrive according to a Poisson process. Each request is for content~$i$ with probability $\pi_i$ independently from previous requests. The catalogue includes 10 thousand contents, divided in two equally-sized classes. Contents in the first class have constant probabilities $\pi_i$ following a Zipf's law with exponent~$0.8$. The same holds for the contents in the second class during one epoch, but at each epoch the probabilities are permuted uniformly at random. 

Two request traces have been collected from two different vantage points of the Akamai CDN. The first one spans over 5 days with $4\times 10^8$ requests for $13\times 10^6$ unique contents. The second one spans over 30 days with $2\times 10^9$ requests for $113\times 10^6$ unique contents.
Detailed information about these traces is available in~\cite{neglia17tompecs}. Both for synthetic and real traces, we considered the contents of equal size (even if information about the real size is available).

All caching policies tested have been implemented in an ad-hoc Python simulator using the PyTorch library for feedforward and LSTM networks. The values of all parameters are those in Table~\ref{t:notation}.

\subsection{Popularity estimation}
\begin{figure}[t]
	\centering
	
	\begin{subfigure}[b]{0.49\linewidth}
		\includegraphics[width=\linewidth]{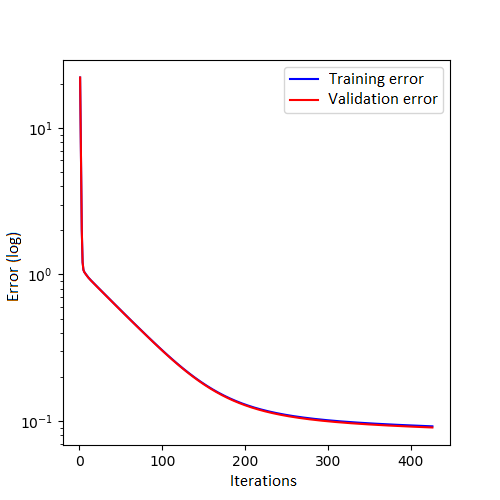}
		\caption{}
	\end{subfigure}
	\begin{subfigure}[b]{0.49\linewidth}
		\includegraphics[width=\linewidth]{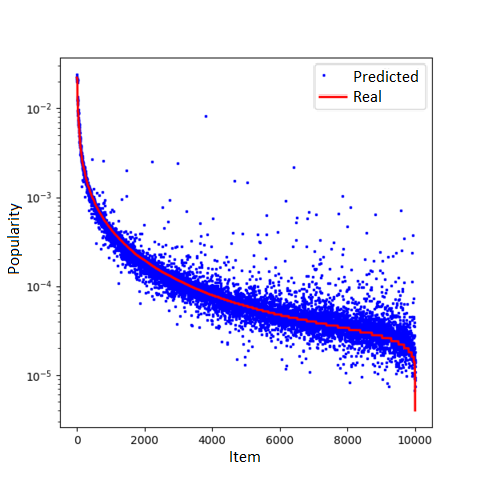}
		\caption{}
	\end{subfigure}
	\caption{Popularity estimation: (a) training and validation errors over the epochs, (b) predicted vs real popularities. Synthetic trace.}
	\label{f:popularity_estimation}
\end{figure}

We first evaluate FNN ability to predict content popularities.
Figure~\ref{f:popularity_estimation} shows the training and validation error versus the number of iterations, where one iteration corresponds to processing 8 requests. The FNN learns quite fast and after about 400 iterations (i.e.~3200 requests), both the training and the evaluation errors reach a plateau (left plot). At this stage, the FNN predicts reasonably well content popularities (right plot). In the synthetic trace there are about $2 \times 10^5$ requests in one epoch, so that at the end of the first epoch the FNN has already learned.
With real traces the traffic characteristics keeps changing and the system keeps learning, but results are qualitatively similar.

We compare the prediction quality for three different predictors: 1)  FNN, 2)  empirical average over the past $K$ popularities (AVG), and 3)  linear regression estimator over the $K$ past popularities (LR).\footnote{
	The linear regression model has been obtained at runtime training an NN without non-linearities (we removed the activation functions).
} Table~\ref{t:prediction_error} shows the MSE for the transformed popularities for the three predictors. 
Both for the synthetic and Akamai 30-day  trace FNN significantly outperforms the other two.

\begin{table}
\centering
\begin{tabular}{| l | c  | c | c |}
\hline
			& FNN	& LR		& AVG\\
\hline
synthetic traces	 & 0.088	& 0.103 &	 0.513 \\
Akamai 30-day trace	 & 0.136	& 0.359	& 0.435 \\
\hline
\end{tabular}
\caption{Prediction MSE for the three predictors.}
\label{t:prediction_error}
\end{table}

\subsection{Comparison with other caching policies}
We compare  our FNN-based caching policy with two popular eviction policies: LRU and ARC.

LRU maintains a priority queue ordered according to the content last access time. Upon a hit, the content is moved to the front of the queue. Upon a miss, the new content is inserted in the cache and the Least Recently Used one is evicted from it. LRU has low-complexity and does not require any configuration. It has shown good performance across a variety of request patterns. For this reason it is one of the most used caching policies. 

Adaptive Replacement Cache (ARC) \cite{3} is a caching policy introduced by IBM in 2004. It uses two priority queues to combine  information about how recently a content has been required (like LRU does) and how frequently (like another policy, LFU, does). The size of the two queues is dynamically adapted to the specific request pattern. ARC is slightly more complex than LRU, but, in our experience, it is very difficult to beat.

We wanted also to compare FNN-caching with more complex NN-based policies. We have considered the DLSTM scheme presented in~\cite{pang18}. We were not able to get the code or the traces from the authors. So, we have implemented it in the same simulator and tried to select all parameters according to the suggestions in~\cite{pang18}.

\begin{figure}[t]
	\centering
	
	\begin{subfigure}[b]{0.49\linewidth}
		\includegraphics[width=\linewidth]{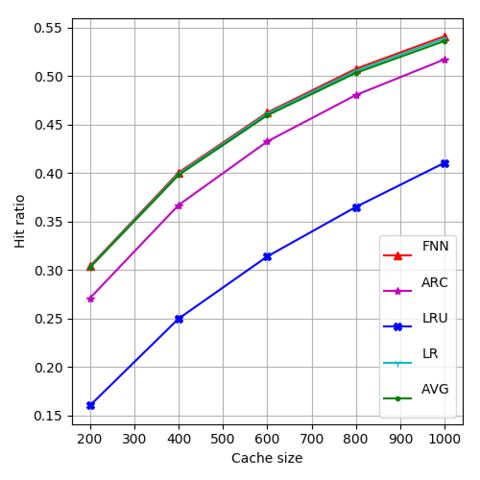}
		\caption{}
		\label{f:caching_a}
	\end{subfigure}
	\begin{subfigure}[b]{0.49\linewidth}
		\includegraphics[width=\linewidth]{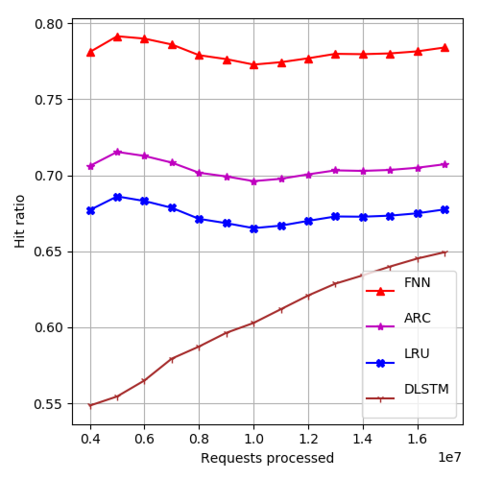}
		\caption{}
		\label{f:caching_b}
	\end{subfigure}
	\caption{ (a) Hit rate for different cache sizes, Akamai 5-day trace. (b) Cumulative hit rate vs number of requests, Akamai filtered 5-day trace (only 1000 contents).}
	\label{f:caching}
\end{figure}


Figure~\ref{f:caching_a} shows the hit rate the different policies achieve over Akamai 5-day trace.  FNN-caching performs significantly better than ARC and LRU. DLSTM does not appear in the figure, because its hit rate was constantly below 1\%. We suspected the poor performance of DLSTM was due to its large number of parameters. In fact, the one-hot encoding used by DLSTM requires a number of weights that grows linearly with the number of contents in the catalogue. As an example, for $10^5$ contents the DLSTM needs to learn more than $10^6$ weights. Our solution instead has about $10^4$ weights independently from the number of contents.  To test this hypothesis, we filtered the trace considering only the requests for the first $1000$ contents appearing in it. Figure~\ref{f:caching_b} supports our hypothesis: the performance of DLSTM keeps improving over time, but, after more than one day, it still lags behind the other policies. Over the 5 days (results not shown here), DLSTM manages to outperform LRU, but neither ARC nor our policy. And, on a larger catalogue, DLSTM is definitely too slow to be competitive.


FNN-caching achieves a higher hit ratio than classic policies like LRU and ARC as well as the new DLSTM. The same conclusion  holds also for the 30-day trace and the synthetic one. But, as the attentive reader may have already remarked, this is not the end of the story. In fact, we have also tested some caching policies where the FNN popularity predictor is replaced by the LR predictor or by the AVG predictor.\footnote{
	The two resulting caching policies belong to the family of Least Recently/Frequently Used (LRFU) policies where the requests numbers during past epochs are weighted with different coefficients~\cite{lee01}.
} The corresponding curves are also in Fig.~\ref{f:caching_a} and are almost indistinguishable from the one of FNN-caching! The hit rate is at most 1\% less for AVG-caching than for FNN-caching and LR-caching has intermediate performance. Hence, while the FNN predictor is significantly better than the other two (see Table~\ref{t:prediction_error}), for caching decision the less precise estimates of LR and AVG are equally good. Moreover, LR has much less parameters to learn ($K+2$) and both LR and AVG are computationally less expensive than FNN.

\section{Discussion and Conclusions}
\label{s:discussion}
Our current results suggest that, for caching purposes, neural network predictors do not have an edge on simpler linear estimators. Moreover, experiments not shown here indicate that considering additional information as input, like the time of the day or the size of the content,  does not improve the performance of FNN-caching.

At the same time, in this paper we assume that past popularities of all contents are known. In reality, this information may be available only for a subset of the contents or may have been lossy compressed (as in~\cite{li16}). It is possible that an NN would work better than a linear predictor with incomplete/noisy inputs.

Moreover, while the LSTM-NN we tested performed poorly, \cite{hashemi18} shows that it may be more advantageous to frame the prediction problem as a classification  rather than as a regression  (as done in all other works~\cite{pang18,zeng17,narayanan18}), because the extreme variability of content popularities ``means that the effective vocabulary size can actually be manageable for RNN [LSTM] models.''

We plan to explore these directions in our future work.

\bibliography{references}

\begin{thebibliography}{10}

\bibitem{chrobak12}
M.~Chrobak, G.~J. Woeginger, K.~Makino, and H.~Xu, ``Caching is hard---even in
  the fault model,'' {\em Algorithmica}, vol.~63, pp.~781--794, Aug 2012.

\bibitem{bastug15}
E.~Ba\c{s}tu\u{g}, M.~Bennis, E.~Zeydan, M.~A. Kader, I.~A. Karatepe, A.~S. Er,
  and M.~Debbah, ``Big data meets telcos: A proactive caching perspective,''
  {\em Journal of Communications and Networks}, vol.~17, pp.~549--557, Dec
  2015.

\bibitem{zeng17}
Y.~Zeng and X.~Guo, ``Long short term memory based hardware prefetcher: A case
  study,'' in {\em Proceedings of the International Symposium on Memory
  Systems}, MEMSYS '17, (New York, NY, USA), pp.~305--311, ACM, 2017.

\bibitem{hashemi18}
M.~Hashemi, K.~Swersky, J.~A. Smith, G.~Ayers, H.~Litz, J.~Chang, C.~E.
  Kozyrakis, and P.~Ranganathan, ``Learning memory access patterns,'' in {\em
  Proc. of the International Conference on Machine Learning (ICML)}, 2018.

\bibitem{tsai18}
K.~C. Tsai, L.~L. Wang, and Z.~Han, ``Caching for mobile social networks with
  deep learning: Twitter analysis for 2016 u.s. election,'' {\em IEEE
  Transactions on Network Science and Engineering}, pp.~1--1, 2018.

\bibitem{narayanan18}
A.~Narayanan, S.~Verma, E.~Ramadan, P.~Babaie, and Z.-L. Zhang, ``Deepcache: A
  deep learning based framework for content caching,'' in {\em Proceedings of
  the 2018 Workshop on Network Meets AI \& ML}, NetAI'18, (New York, NY, USA),
  pp.~48--53, ACM, 2018.

\bibitem{zhang18}
N.~Zhang, K.~Zheng, and M.~Tao, ``Using grouped linear prediction and
  accelerated reinforcement learning for online content caching,'' in {\em 2018
  IEEE International Conference on Communications Workshops (ICC Workshops)},
  pp.~1--6, May 2018.

\bibitem{pang18}
H.~Pang, J.~Liu, X.~Fan, and L.~Sun, ``Toward smart and cooperative edge
  caching for 5g networks: A deep learning based approach,'' in {\em Proc. of
  IEEE/ACM International Symposium on Quality of Service (IWQoS)}, 2018.

\bibitem{zhong18}
C.~Zhong, M.~C. Gursoy, and S.~Velipasalar, ``A deep reinforcement
  learning-based framework for content caching,'' in {\em Proc. of the 52nd
  Annual Conference on Information Sciences and Systems (CISS)}, 2018.

\bibitem{rezaei18}
E.~Rezaei, H.~E. Manoochehri, and B.~H. Khalaj, ``Multi-agent learning for
  cooperative large-scale caching networks,'' {\em ArXiv e-prints}, 2018.
\newblock arXiv:1807.00207.

\bibitem{henderson18}
P.~Henderson, R.~Islam, P.~Bachman, J.~Pineau, D.~Precup, and D.~Meger, ``Deep
  reinforcement learning that matters,'' {\em ArXiv e-prints}, 2017.
\newblock arXiv:1709.06560.

\bibitem{li16}
S.~Li, J.~Xu, M.~van~der Schaar, and W.~Li, ``Popularity-driven content
  caching,'' in {\em Proc. of the 35th Annual IEEE International Conference on
  Computer Communications (INFOCOM)}, 2016.

\bibitem{14}
M.~M. Lau and K.~H. Lim, ``Investigation of activation functions in deep belief
  network,'' in {\em Proc. of the 2nd International Conference on Control and
  Robotics Engineering (ICCRE)}, 2017.

\bibitem{16}
R.~M. French, ``Catastrophic forgetting in connectionist networks,'' {\em
  Trends in Cognitive Sciences}, vol.~3, no.~4, pp.~128--135, 1999.

\bibitem{17}
J.~Kirkpatricka, R.~Pascanua, {\em et~al.}, ``Overcoming catastrophic
  forgetting in neural networks,'' {\em National Academy of Sciences}, 2017.

\bibitem{neglia17tompecs}
G.~Neglia, D.~Carra, M.~Feng, V.~Janardhan, P.~Michiardi, and D.~Tsigkari,
  ``Access-time-aware cache algorithms,'' {\em ACM Trans. Model. Perform. Eval.
  Comput. Syst.}, vol.~2, pp.~21:1--21:29, Nov. 2017.

\bibitem{3}
N.~Megiddo and D.~S. Modha, ``Outperforming lru with an adaptive replacement
  cache algorithm,'' {\em Computer}, vol.~37, no.~4, pp.~58--65, 2004.

\bibitem{lee01}
D.~Lee, J.~Choi, J.~H. Kim, S.~H. Noh, S.~L. Min, Y.~Cho, and C.~S. Kim,
  ``Lrfu: A spectrum of policies that subsumes the least recently used and
  least frequently used policies,'' {\em IEEE Trans. Comput.}, vol.~50,
  pp.~1352--1361, Dec. 2001.

\end{thebibliography}
\bibliographystyle{ieeetr}

\end{document}